\documentclass[a4paper,10pt]{article}

\usepackage{cite}
\usepackage{color}
\usepackage{graphicx}
\usepackage[usenames,dvipsnames]{xcolor} 

\definecolor{DarkGreen}{RGB}{0,64,0}

\usepackage[]{array}
\usepackage[]{amsmath}
\usepackage[]{amssymb}
\usepackage[]{mathrsfs}
\usepackage[]{tensor}
\usepackage{bm}

\begin{document}

\vspace*{2cm}

\begin{center}

{\LARGE{\bf Field redefinitions, Weyl invariance, and\\[2mm] nature of mavericks}}
\vskip 1cm

{\Large Predrag Dominis Prester}\\
{}~\\
\quad \\
{\em Department of Physics, University of Rijeka,}\\[1mm]
{\em  Radmile Matej\v{c}i\'{c} 2, HR-51000 Rijeka, Croatia}\\
\vskip 1cm
Email: pprester@phy.uniri.hr

\end{center}

\vskip 1.5cm 

\noindent
{\bf Abstract.}
In the theories of gravity with non-minimally coupled scalar fields there are ``mavericks'' -- unexpected solutions with odd properties, e.g., black holes with scalar hair in theories with scalar potential bounded from below. Probably the most famous example is Bocharova-Bronnikov-Melnikov-Bekenstein (BBMB) black hole solution in a theory with a scalar field conformally coupled to the gravity and with vanishing potential. Its existence naively violates no-hair conjecture without violating no-hair theorems because of the singular behavior of the scalar field at the horizon. Despite being discovered more than 40 years ago, nature of BBMB solution is still the subject of research and debate. We argue here that the key in understanding nature of maverick solutions is the proper choice of field redefinition schemes in which the solutions are regular. It appears that in such ``regular'' schemes mavericks have different physical interpretations, in particular they are not elementary but composite objects. For example, BBMB solution is not an extremal black hole but a collection of a wormhole and a naked singularity. In the process we show that Weyl-invariant formulation of gravity is a perfect tool for such analyzes.
\medskip

\noindent

\vskip 1cm 

 
\vfill\eject

\section{Introduction}
\label{sec:intro}

\medskip

Recent developments in cosmology and elementary particle physics, highlighted by the discovery of Higgs boson and the results from Planck and BICEP experiments, imply that it is imperative to improve our understanding of mechanisms for coupling scalar fields to gravity. As a consequence, there has been a revived interest recently in theories with scalar fields nonminimally coupled to gravity. An important and well-known aspect of such theories is that some of them allow new types of solutions which do not have counterparts in standard general relativity with minimally coupled scalars, even in cases in which exist locally well-defined field redefinitions which transform the theory into general relativity with minimally coupled scalars. An interesting example of such solutions are ``maverick'' black holes. 

To be specific, let us consider the theory with a scalar field $\chi(x)$ nonminimally coupled to gravity with the following ``Jordan frame'' action 
\begin{equation} \label{ccsc}
I_c = \int d^4 x \sqrt{-g} \left[ \frac{1}{2\kappa} (R - 2 \Lambda) - \frac{1}{2}(\partial \chi)^2
 - \frac{1}{12} \chi^2 R - V_c(\chi) \right]
\end{equation}
where $\kappa = 8\pi G_N$  ($G_N$ being Newton constant) and $\Lambda$ is the cosmological constant. For the class of theories\footnote{Theory (\ref{ccsc}) with scalar potential of the form $V_c = \lambda \chi^4$ are sometimes referred to have conformally coupled scalar field, though the gravity part of the action is not conformally invariant. Because of this and also to not confuse a reader when we introduce (full) Weyl invariance later in the text, we shall avoid here using the word conformal in this context.} obtained by taking the scalar potential $V_c(\chi)$
\begin{equation} \label{ccpot}
V_c(\chi) = - \frac{\kappa \Lambda}{36} \chi^4
\end{equation}
maverick black hole solutions have been found. In particular, for the simplest case in which 
$\Lambda=0$ this theory has the following solution \cite{BBM,Bekenstein:1974sf}
\begin{equation}\label{bbmb}
ds^2 = - f(r) dt^2 + \frac{dr^2}{f(r)} + r^2 (d\theta^2 + \sin^2 \theta d\varphi^2) \;,\qquad
 \chi(r) = \pm \sqrt{\frac{6}{\kappa}} \frac{M}{r-M}
\end{equation}
where function $f(r)$ is given by
\begin{equation}\label{bbmbf}
f(r) = \left(1 - \frac{M}{r}\right)^{\!2}
\end{equation}
As the metric in (\ref{bbmb}-\ref{bbmbf}) is the same as for the extremal Reissner-Nordstrom solution, it appears that this solution, usually called Bocharova-Bronnikov-Melnikov-Bekenstein (BBMB) solution, describes asymptotically flat static spherically symmetric black hole with mass $M$, an event horizon at $r=M$ and a timelike space-time singularity at $r=0$. A generalized versions carring $U(1)$ electric and magnetic charge were also constructed \cite{BBM,Bekenstein:1974sf,Virbhadra:1993st}, and a generalization to $\Lambda\ne0$, called MTZ solution, was obtained in \cite{Martinez:2002ru}. BBMB black hole carries (discrete) hair, which means that its existence apparently breaks no-hair conjecture.\footnote{Note that there is also a standard Schwarzschild solution with everywhere vanishing scalar field. For a review and history of no-hair theorems see \cite{Bekenstein:1996pn}.} Both the dominant and strong energy conditions hold for the BBMB solution and its generalizations. 

A tricky feature of BBMB solution is that the scalar field is singular at the horizon.\footnote{For this reason BBMB solution does not violate no-hair theorems. For an analysis of no-hair theorems for theories of the type (\ref{ccsc})-(\ref{ccpot}) see \cite{AyonBeato:2002cm}. As we shall see, for MTZ solution with $\Lambda\ne0$ scalar field singularity does not coincide with the event horizon and is inside the black hole.} As shown in \cite{Bekenstein:1975ts}, this singularity is harmless for classical particle trajectories and tidal accelerations, so it existence does not automatically imply that the solution is pathological at the horizon. However, further analyses revealed some potentially problematic aspects of the interpretation of BBMB solution as a genuine black hole solution. Black hole thermodynamics is not well-defined \cite{Zaslavskii:2002zv}. Continuity of the equations of motion across the horizon is broken \cite{Sudarsky:1997te}. It is not clear what should be the proper boundary condition on the horizon for perturbations, leading to controversy regarding (un)stability of solution under linear spherical perturbations \cite{Bronnikov:1978mx,McFadden:2004ni}. Also, perturbatively constructed simple separable rotating solution diverges at $r=2M$ \cite{Bhattacharya:2013hvm}. One additional strange property of BBMB solution is that in the region $0<r<2M$ effective Newton constant is negative (``antigravity'' behavior).

The continuing interest in BBMB solution gives a strong motive for understanding the real nature of such maverick solutions. We suggest here that a key to achieve this goal is in a proper choice of a field redefinition scheme. We argue that the singularity of scalar field is a consequence of a breakdown of Jordan frame scheme at $r=M$, and show that in schemes which are more ``natural'' (i.e., regular) for describing such solutions, maverick solutions have a different physical interpretation, e.g., BBMB (or MZT) solution is a collection of two objects, a wormhole solution and a naked singularity, separated by an asymptotic region located around $r  = M$. We demonstrate that for analyzes of this type it is very convenient to use Weyl-invariant dilaton formulation of gravity, which in some cases allows easy understanding of limits of particular schemes and a construction of field redefinition schemes which are more convenient for a given purpose or configuration.

\vspace{10pt}

\section{Weyl-invariant dilaton gravity}
\label{sec:wysm}

\medskip

\begin{figure}[htb]
\begin{center}
\includegraphics[scale=0.8]{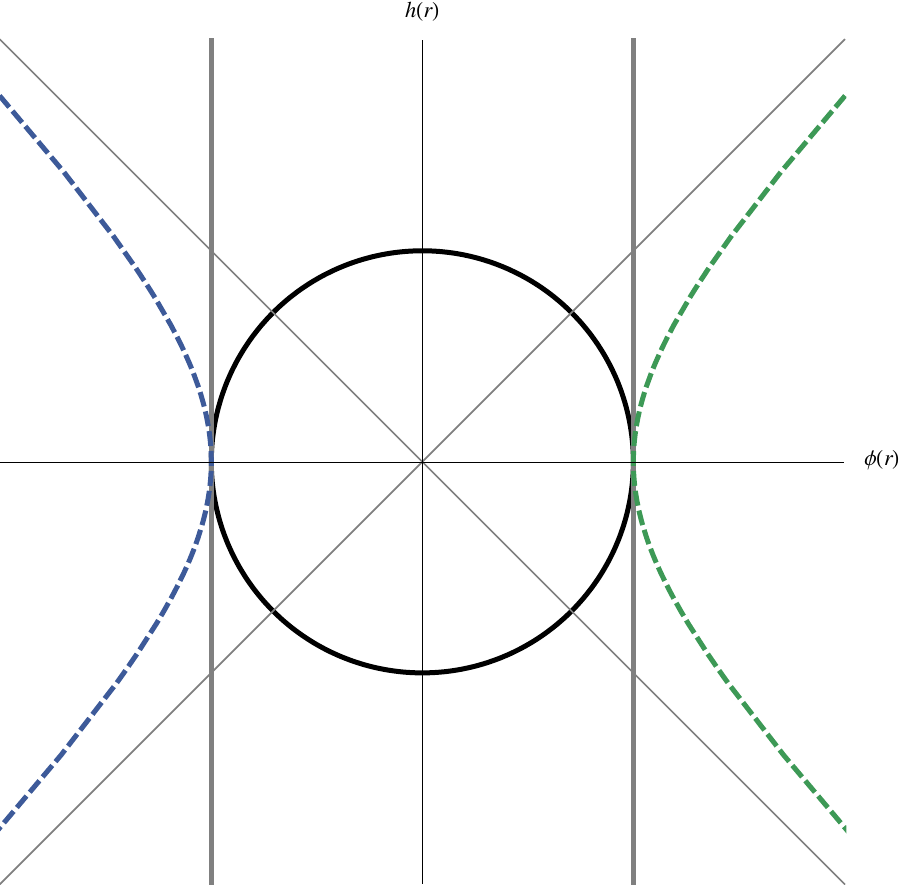}
\caption{We plot curves in $\phi$--$h$ plane defined by gauge fixing condition (\ref{gfix}) for three gauges discussed here: E-gauge (dashed lines, hyperbola), c-gauge (thick gray vertical straight lines), and k-gauge (thick black lines, circle). The grey thin straight lines are $|\phi| = |h|$ graphs, which divide the plane into four wedges. Due to the Weyl-invariance all points on a given radial line (excluding the origin $(0,0)$) are equivalent. We see that E-gauge can accommodate only configurations which lie exclusively in either left or right wedge (without crossing from one to another). As for c-gauge, it is not defined on $\phi=0$ line, so all configurations it accommodates lie exclusively either in left or right half-plane (they cannot cross $\phi=0$ line). As for k-gauge, it obviously covers the whole plane, except for the point $(0,0)$. Configurations which in k-gauge (or any other gauge which regularly accommodates $\phi=0$ lines) have $\phi_k(x)=0$ at spacetime point $x$, would in c-gauge give $|h_c(x)|\to\infty$. We shall argue that this is a true nature of singularities present in BBMB and MTZ solutions at $r=M$.}
\label{fig:gauges}
\end{center}
\end{figure}

Here we briefly review a Weyl-invariant formulation of dilaton gravity (WIDG), with the main idea to embed the theory (\ref{ccsc}) into this formalism. Our interest here is a gravity theory with matter sector consisting of one (physical) scalar field $h(x)$.\footnote{WIDG can naturally encompass much more complicated theories, including Standard model of particle physics with minimal or nonminimal Higgs sector. For this, and more detailed review of WIDG, one can consult \cite{Bars:2013yba}.} In this case one can write a manifestly Weyl-invariant actions by including an additional scalar field $\phi(x)$
\begin{equation} \label{Iwism}
I_{\mathrm{WIDG}} = \int d^4 x \sqrt{-g} \left[ \frac{1}{12} (\phi^2 - h^2) R 
 + \frac{1}{2} (\partial \phi)^2 - \frac{1}{2} (\partial h)^2 - V(\phi,h) \right] \;,\qquad  V(\phi,h) = \phi^4\, F(h/\phi)
\end{equation}
This action is invariant on Weyl transformations defined by
\begin{eqnarray} \label{weyltr}
g^{\mu\nu}(x) \,\to\, \Omega(x)^2\, g^{\mu\nu}(x) \;,\qquad 
\phi(x) \,\to\, \Omega(x)\, \phi(x) \;,\qquad  h(x) \,\to\, \Omega(x)\, h(x)
\end{eqnarray}
where $\Omega(x) \ne 0$ is an otherwise arbitrary function. Weyl invariance by itself does not constrain the function $F$ in (\ref{Iwism}), however if we require for the scalar potential $V$ to be regular in the whole $\phi$-$h$ plane then it has to be of the form $V(\phi,h)=\sum_{n=0}^4 c_n h^n \phi^{4-n}$ where $c_n$ are some coupling constants (or, in other words, $F$ should be a polynomial maximally of the order four). If one regards Weyl rescalings as gauge transformations, one scalar field can be gauged away and we are left with the same number of degrees of freedom we started with (i.e., as in (\ref{ccsc})). We refer to this theory as Weyl invariant dilaton gravity (WIDG). As is standard in gauge theories we can deal with Weyl invariance by fixing the gauge. A convenient type of gauge fixing conditions, which keep manifest diff-covariance and 2-derivative nature of the action, is
\begin{equation} \label{gfix}
f(\phi,h) = 0
\end{equation}
where $f$ is some function defining a gauge. Obviously, a gauge fixing condition of the type (\ref{gfix}) defines a curve in $\phi$--$h$ plane. As $h/\phi$ is gauge invariant, gauge orbits in $\phi$--$h$ plane are radial straight lines plus $(0,0)$ point. It then follows that a gauge will be well-defined in some region of $\phi$--$h$ plane if the curve defined by its gauge condition (\ref{gfix}) crosses every orbit from the region once. We now specify three different gauge choices, visualized in Fig. \ref{fig:gauges}, which we shall refer to later.

\medskip

\noindent
\underline{\emph{Einstein gauge (E-gauge):}} The gauge-fixing condition is
\begin{equation} \label{Egauge}
\phi_E^2 - h_E^2 = \frac{6}{\kappa}
\end{equation}
In this gauge WIDG action (\ref{Iwism}) takes the form of the standard general relativity, with Newton constant $G_N = \kappa/8\pi$, and with one minimally coupled scalar field $\sigma$ and potential $V_E$ defined by
\begin{equation} \label{mcsc}
\sigma = \sqrt{\frac{6}{\kappa}} \tanh^{-1} \left( \frac{h}{\phi} \right) \;, \qquad
V_E(\sigma) = \frac{\kappa^2}{36} \cosh^4 \!\left( \sqrt{\frac{\kappa}{6}} \sigma \right) 
 F\!\left( \tanh \left( \sqrt{\frac{\kappa}{6}} \sigma \right) \right) \;.
\end{equation}
The gauge-fixing condition (\ref{Egauge}) is obviously singular on the lines $|\phi| = |h|$, and as a consequence is the (gauge invariant) field $\sigma$. It then follows that every configuration which is regular in E-gauge must be completely contained inside one of the wedges defined by $|h|<|\phi|$ (right and left wedge in Fig. \ref{fig:gauges}). In conclusion, we see that WIDG theory is equivalent to ordinary general relativity (with scalar field(s)) in the subset of its configuration space.

\medskip

\noindent
\underline{\emph{Jordan gauge (c-gauge):}} The gauge-fixing condition is
\begin{equation} \label{cgauge}
\phi_c^2 = \frac{6}{\kappa}
\end{equation}
After defining $\chi(x) \equiv h_c(x)$ and $V_c(\chi) \equiv V(\phi_c,h_c) - \Lambda/\kappa$ action in c-gauge becomes
\begin{equation} \label{cgI}
I_c = \int d^4 x \sqrt{-g_c}
 \left[ \frac{1}{2\kappa} (R_c - 2 \Lambda) - \frac{1}{2}(\partial \chi)^2 - \frac{1}{12} \chi^2 R_c - V_c(\chi) \right]
\end{equation}
where subscript $c$ on Ricci scalar $R$ denotes that it is computed from the metric in c-gauge ($g_{c\mu\nu}$). We see that in 
c-gauge WIDG action (\ref{Iwism}) takes the form of the action (\ref{ccsc}). If we specify WIDG scalar potential to the form
\begin{equation} \label{vcubic}
V(\phi,h) =  \frac{\kappa \Lambda}{36} (\phi^4 - h^4)
\end{equation}
then in c-gauge WIDG action (\ref{Iwism}) takes \emph{exactly} the form of the action (\ref{ccsc})-(\ref{ccpot}). 

The gauge fixing condition (\ref{cgauge}) is singular on the line $\phi=0$, so it follows that every configuration which is regular in c-gauge is confined to one of the half-planes $\phi>0$ or $\phi<0$. Obviously, the space of regular configurations in the conformal frame is larger then in the Einstein frame. In particular, it allows configurations which cross $|h|=|\phi|$ line. This observation explains why there are solutions in c-gauge, such as maverick black holes, which do not have counterparts in E-gauge. Note that such solutions are necessarily characterized by the property that in the part of the spacetime effective Newton constant $G_\mathrm{eff} = 3/4\pi(\phi^2-h^2)$ is negative (``antigravity region'').\footnote{Note that effective Newton constant $G_\mathrm{eff}$ is singular on the lines $|\phi|=|h|$, which possibly may bring problems to solutions crossing these lines (such as instability on small perturbations).} But, if a configuration crosses $\phi = 0$ line, in c-gauge this would manifest as a singular behavior because $|h_c| \to \infty$ when $\phi \to 0$.

\medskip

\noindent
\underline{\emph{k-gauge}:} The gauge-fixing condition is
\begin{equation} \label{kgauge}
\phi_k^2 + h_k^2 = \frac{6}{\kappa}
\end{equation}
This gauge is regular on the whole $(\phi,h)$ plane except at the Weyl-invariant point $(0,0)$, so in this sense it is superior to both E- and c-gauges. In particular, it can be used to describe solutions crossing $\phi=0$ (with $h\ne0$) lines, which is the propery we shall require later. One convenient parametrization is to introduce the field $\beta(x)$ through
\begin{equation} \label{kgfield}
\phi_k = \sqrt{\frac{6}{\kappa}} \cos \beta \;, \qquad
h_k = \sqrt{\frac{6}{\kappa}} \sin \beta \;, \qquad
\beta = \arctan \frac{h}{\phi}
\end{equation}
$\beta$ is the (gauge invariant) polar angle in $\phi$--$h$ plane, and so it is periodic with period $2\pi$. After gauge fixing the WIDG action becomes
\begin{equation} \label{kgL}
I_k = \frac{1}{2\kappa} \int d^4 x \sqrt{-g_k} \cos(2\beta) \left[ R_k - 6 (\partial \beta)^2 - V_k(\beta) \right]
\;,\qquad
V_k(\beta) = \frac{2\kappa}{\cos(2\beta)} V(\phi_k(\beta),h_k(\beta))
\end{equation}
For the choice of the WIDG potential given in (\ref{vcubic}) one obtains
\begin{equation} \label{vkcubic}
V_k = 2\Lambda
\end{equation}
i.e., the theory in k-gauge describes a sine-Gordon scalar with noncanonical kinetic term. 

\smallskip

\begin{center} * * * \end{center}

\smallskip

Let us make here two observations.

The first observation is that actions corresponding to different gauge choices are related by field redefinitions. In this language, E-gauge and c-gauge correspond to well-known schemes usually called Einstein frame and Jordan frame, respectively. As for k-gauge, we are not aware that corresponding scheme was studied before, but we have shown that it follows very naturally from WIDG framework. A connection between any two gauges, denoted by A and B, is of the form
\begin{equation} \label{gtrans}
g_A^{\mu\nu} \,=\, \Omega_{AB}^2\, g_B^{\mu\nu} \;,\qquad 
\phi_A \,=\, \Omega_{AB}\, \phi_B \;,\qquad  h_A \,=\, \Omega_{AB}\, h_B
\end{equation}
It is easy to show that the factors $\Omega$ for transitions from c-gauge to E-gauge and 
k-gauge are given, respectively, by
\begin{equation} \label{transom}
\Omega_{Ec} = \left( 1 - \frac{\kappa}{6} h_c^2 \right)^{-\frac{1}{2}} \;, \qquad
\Omega_{kc} = \left( 1 + \frac{\kappa}{6} h_c^2 \right)^{-\frac{1}{2}}
\end{equation}
We see that WIDG formulation offers a convenient practical tool for dealing with field redefinition schemes through: (i) easy construction of connecting relations between schemes by using Eqs. (\ref{weyltr}) and (\ref{gfix}), (ii) providing a domain of configuration space covered by schemes, (iii) constructing convenient schemes for particular purposes.

The second observation is that domains in configurations space may also be limited by the possible singularities of the scalar potential $V(\phi,h)$. As an example consider a theory in E-gauge (Einstein frame). First note that Weyl invariance by itself does not restrict a form of the function $F$ in (\ref{Iwism}), and as a consequence potential in Einstein frame $V_E(\sigma)$ is also arbitrary. However, if one requires for WIDG potential $V(\phi,h)$ to be regular in the whole $\phi$--$h$ plane then $F$ must be a polynomial of (maximal) order 4, which then severely restricts $V_E(\sigma)$. If we want to have a standard polynomial potential in Einstein frame
\begin{equation}
V_E(\sigma) = \sum_{j=0}^k \lambda_j \, \sigma^j \;,
\end{equation}
where $\lambda_j$ are some coupling constants, which looks ``natural'' from Einstein frame perspective, then from (\ref{mcsc}) follows that corresponding WIDG potential $V(\phi,h)$ is singular at $|\phi| = |h|$. This may prevent existence of regular configurations crossing these lines in other gauges also (note that configuration space in Einstein frame is limited by these lines by definition). Another example of a singular potential is obtained by allowing $F$ to be polynomial of an order higher then 4. In this case potential is singular on the line $\phi=0$, which can put a boundary on regular solutions in general gauges (configuration space in Jordan frame is by definition limited by $\phi=0$ line). The similar argumentation applies to couplings of the scalar fields to the matter sector, when such exist (we ignore them in this paper). In the rest of the paper we analyze solutions in theories with WIDG potentials regular in the whole $\phi$--$h$ plane, so this issue will not appear in our analyses here.

\vspace{10pt}

\section{BBMB solution}
\label{sec:bbmb}

\medskip

We now turn our attention to BBMB solution (\ref{bbmb})-(\ref{bbmbf}). As the metric is the same as for extremal Reisner-Nordstom solution with mass $M$ it apparently describes spherically symmetric asymptotically flat extremal black hole with event horizon located at $r=M$. BBMB black hole is a solution of the action (\ref{ccsc})-(\ref{ccpot}) with $\Lambda=0$. We have shown in the previous section that this theory may be described as WIDG (\ref{Iwism}) with the potential $V=0$ written in c-gauge. In this language BBMB solution is
\begin{eqnarray} \label{bbmbsg}
&& ds_c^2 = - f(r) dt^2 + \frac{dr^2}{f(r)} + r^2 (d\theta^2 + \sin^2 \theta d\varphi^2) \;,\qquad
f(r) = \left( 1 - \frac{M}{r} \right)^{\!\!2}
\nonumber \\
&& \phi_c(r) = \sqrt{\frac{6}{\kappa}} \;,\qquad
h_c(r) = \sqrt{\frac{6}{\kappa}}\, \frac{M}{r-M}
\end{eqnarray}
The ``ugly'' property of BBMB solution is the singular behavior of the scalar field $h_c$ at $r=M$. 

At $r=2M$ one has $\phi = h$, while for $r\to 0$ one gets $|h| \to |\phi|$. It follows that in the region $r<2M$ one has $|\phi| < |h|$. As discussed in Sec. \ref{sec:wysm}, E-gauge cannot accommodate such antigravity behavior and this immediately explains why there is no a corresponding regular solution in Einstein frame covering $r>M$ region.

The more intriguing aspect of BBMB solution is the singular behavior of the field $h_c$ at $r=M$. Using WIDG picture (and Fig. \ref{fig:gauges}) it is easy to understand the nature of this singularity -- as $h_c \to \infty$ this signals that c-gauge breaks down, which can only happen when 
$\phi \to 0$ in regular gauges (i.e., regular field redefinition schemes). There are essentially two possible scenarios: (a) $h\ne0$ at $r=M$, (b) $h=0$ at $r=M$.

Let us envisage the scenario (a) first. In this case we can use field redefinition scheme corresponding to k-gauge (\ref{kgauge}), as it implies $h_k^2 = 6/\kappa > 0$ for $\phi_k = 0$. An action in k-gauge is by (\ref{kgL})
\begin{equation}
I_k = \frac{1}{2\kappa} \int d^4 x \sqrt{-g_k} \cos(2\beta) \left[ R_k - 6 (\partial \beta)^2 \right]
\end{equation}
We do not have to solve corresponding field equations, as the transition rule from c-gauge to k-gauge is already given in (\ref{gtrans}) and second equation of (\ref{transom}). Inserting (\ref{bbmbsg}) into (\ref{gtrans})-(\ref{transom}) we obtain the expression for BBMB solution in k-gauge
\begin{equation} \label{bbmbkg}
g_k^{\mu\nu} \,=\, \Omega_{kc}^2\, g_c^{\mu\nu} \;,\qquad 
\phi_k \,=\, \Omega_{kc}\, \phi_c \;,\qquad  h_k \,=\, \Omega_{kc}\, h_c \;,\qquad
\Omega_{kc} \,=\, (r-M) \left[ (r-M)^2 + M^2 \right]^{-\frac{1}{2}}
\end{equation}
Explicitely written, BBMB solution in k-gauge scheme is
\begin{eqnarray} \label{bbmbkge}
&& ds_k^2 \,=\, \left[ \left(1-\frac{M}{r}\right)^{\!\!2} + \frac{M^2}{r^2} \right]
 \left( - dt^2 + \frac{dr^2}{\left( 1 - \frac{M}{r} \right)^{\!4}}
 + \frac{r^2}{\left( 1 - \frac{M}{r} \right)^{\!2}} (d\theta^2 + \sin^2 \theta d\varphi^2) \right)
\nonumber \\
&& \beta(r) \,=\, \arctan \frac{M}{r-M}
\end{eqnarray}
where the``physical'' k-gauge scalar field $\beta$ is obtained from (\ref{kgfield}).

Let us analyze more closely this solution. By construction, scalar field singularity $r=M$ is gone and all k-gauge scalar fields (including $\beta$) are regular in the whole interval $0 \le r < \infty$. The field $\beta$ is 0 at $r\to\infty$, $\pi/2$ at $r=M$, and $3\pi/4$ 
(which means $h=-\phi$) at $r=0$. As the factor $\Omega_{kc}$ behaves as
\begin{equation} \label{okcas}
\Omega_{kc}(r) \,\stackrel{r\to 0}{\longrightarrow}\, \frac{1}{\sqrt{2}} + O(r) \;,\qquad
\Omega_{kc}(r) \,\stackrel{r\to \infty}{\longrightarrow}\, 1 + O(1/r^2)
\end{equation}
the interpretations of the limits $r\to0$ and $r\to\infty$ are essentially unchanged by the field redefinition. What remains is to analyze the metric (\ref{bbmbkge}) in $r \approx M$ region. It can be shown that components of Riemann tensor $R^{\mu\nu}{}_{\rho\sigma}$ all vanish at $r=M$ (as well as all curvature diff-invariants constructed out of metric, Riemann tensor and covariant derivatives). 

To understand the region $r>M$, first note that ``Schwarzschild'' radius $r/\Omega_{kc}(r)$ has the minimum at $r=2M$. It is convenient to pass to the new radial coordinate $\mathcal{R}$ defined by
\begin{equation} \label{newradc}
\mathcal{R}^2 + 8M^2 \,\equiv\, r^2\, \Omega_{kc}(r)^{-2} 
\end{equation}
The coordinate $\mathcal{R}$ is defined to go from $-\infty$ to $+\infty$ as r goes from $M$ to $+\infty$. In the new coordinate system metric in the asymptotic regions $\mathcal{R}\to\pm\infty$ behaves as
\begin{equation} \label{bbmbkgas}
ds_k^2 = - \left(1-\frac{2M}{|\mathcal{R}|}+O(\mathcal{R}^{-2})\right) dt^2
 + \left(1+\frac{2M}{|\mathcal{R}|}+O(\mathcal{R}^{-2})\right) dt^2
 + \left(\mathcal{R}^2+8M^2\right) (d\theta^2 + \sin^2 \theta d\varphi^2) 
\end{equation}
Taking all together, we have established that the patch of spacetime defined by $M<r<\infty$ describes a wormhole with two asymptotically flat infinities ($r\to M$ and $r\to\infty$) connected by the throat centered at $r=2M$ ($\mathcal{R}=0$) and with proper radius equal to $2\sqrt{2}M$. 

As for $0<r<M$ region, its nature is best understood by using as a new (``Schwarzschild'')  radial coordinate $\mathcal{R} \equiv r/\Omega_{kc}$ with a domain $0<\mathcal{R}<\infty$. In the limit $r\to M^-$ ($\mathcal{R}\to\infty$) the metric in k-gauge is again of the form (\ref{bbmbkgas}) except that now $\mathcal{R}^2 + 8M^2 \to \mathcal{R}^2$. We conclude that the metric in the region $0<r<M$ describes an asymptotically flat naked singularity. 

It is not hard to see that these conclusions are not exclusive to k-gauge but are valid in all regular gauges in which scalar fields are regular everywhere and $h\ne0$ at $r=M$. Now we can come back to the scenario (b), i.e., to gauges in which $h=0$ at $r=M$. It is rather easy to see that in such gauges metric is singular at $r=M$, so these are not regular gauges for BBMB solution. A final conclusion is that \emph{in all regular field redefinition schemes BBMB solution does not describe a single black hole but a collection of an asymptotically flat naked singularity and an asymptotically flat wormhole.}

\vspace{10pt}

\section{MTZ solution}
\label{sec:mtz}

\medskip

BBMB solution is a special case of a larger class of solutions of the actions (\ref{ccsc})-(\ref{ccpot}) parametrized by the value of the cosmological constant $\Lambda$ beside the mass parameter 
$M$. This class, known as Martinez-Troncoso-Zanelli (MTZ) solution, is again given by (\ref{bbmb}), but with the function $f$ now being
\begin{equation}\label{mtzf}
f(r) = \left(1 - \frac{M}{r}\right)^{\!\!2} - \frac{\Lambda}{3} r^2
\end{equation}
For $\Lambda=0$ one gets BBMB solution. If $\Lambda\ne0$, then an event horizon is present only if
\begin{equation} \label{mtzbh}
\Lambda > 0 \qquad \mathrm{and} \qquad 0 < M < \frac{\ell}{4}
 \equiv \frac{1}{4} \sqrt{\frac{3}{\Lambda}} \;.
\end{equation}
In this case the metric is the same as for Reissner-Nordstr\"{o}m-deSitter black hole with the special value of charge ($|Q|=M$ in standard conventions).\footnote{For such RNdS black holes temperatures of the event horizon and the cosmological horizon are the same, so they are thermodynamically stable. For detailed analysis see \cite{Brill:1993tw}.} We focus on this case, which we refer to as MTZ solution in the rest of the paper.\footnote{In the rest of the parameter space MTZ solution describes naked singularity which is asymptotically dS or AdS depending on the sign of $\Lambda$. Our analysis can be easily extended to these cases.} The equation $f(r_h)=0$ then has three solutions
\begin{equation} \label{mtzhor}
r_- = \frac{\ell}{2} \left( \sqrt{1 + \frac{4M}{\ell}} - 1 \right) \;,\qquad 
r_+ = \frac{\ell}{2} \left( 1 - \sqrt{1 - \frac{4M}{\ell}} \right) \;,\qquad
r_{++} = \frac{\ell}{2} \left( 1 + \sqrt{1 - \frac{4M}{\ell}} \right)
\end{equation}
satisfying 
\begin{equation}
0 < r_- < M < r_+ < 2M < \frac{\ell}{2} < r_{++} < \ell
\end{equation}
where $r_+$ is the outer (event) horizon, $r_-$ is the inner (Cauchy) horizon and $r_{++}$ is the cosmological horizon. There are important differences between BBMB and MTZ solutions. MTZ solution is non-extremal and asymptotically dS. In addition, the singularity of scalar field, located at $r=M$, is inside the MTZ solution so the event horizon is completely regular.\footnote{The reason why MTZ solution does not violate no-hair theorems is that the scalar potential (\ref{ccpot}) is not bounded from below for $\Lambda > 0$. There are many known black hole solutions with scalar hair in theories with potential unbounded from below (including those in standard GR with minimally coupled scalar fields) which are completely regular except for the central singularity. So, strictly speaking, one could say that MTZ is not a maverick solution.} Our aim here is to show that despite these differences the analysis applied to BBMB solution can be straightforwardly repeated here.

We start by observing that singularity of the scalar field at $r=M$ appears because field redefinition scheme breaks there. In WIDG language this happens because $\phi=0$ at $r=M$, and this cannot be regularly represented in c-gauge. The gauge which does not have the problem with $\phi=0$ (if $h\ne0$) is k-gauge (\ref{kgauge}) so we can use it to obtain regular description of MTZ solution. The Eqs. (\ref{bbmbsg})-(\ref{okcas}) with the corresponding consequences apply again here, with the only difference that the function $f(r)$ is now given in (\ref{mtzf}). MTZ solution in a field redefinition scheme corresponding to k-gauge (in which action is given in 
(\ref{kgL})-(\ref{vkcubic})) is then given by
\begin{eqnarray} \label{mtzkg}
&& ds_k^2 \,=\, \frac{(r-M)^{\!2} + M^2}{(r-M)^2}
 \left\{ - \left[(r-M)^{2} - \frac{\Lambda}{3} r^4 \right] \frac{dt^2}{r^2} + \frac{r^2\, dr^2}{(r-M)^2 - \frac{\Lambda}{3} r^4}
 + r^2 (d\theta^2 + \sin^2 \theta d\varphi^2) \right\}
\nonumber \\
&& \beta(r) \,=\, \arctan \frac{M}{r-M}
\end{eqnarray}
Let us analyze this solution. Again, $r\to0$ and $r\to\infty$ behavior is unchanged by passing to k-gauge. However, a behavior of the metric near $r=M$ is different than in c-gauge.  

For $r>M$ it is convenient to switch to a new coordinate $\mathcal{R}$ defined in (\ref{newradc}), which goes from $-\infty$ to $\infty$ as $r$ goes from $M$ to $\infty$. The metric in k-gauge for $r\to M$ ($\mathcal{R}\to -\infty$) is then approximately given by 
\begin{equation} \label{mtzkgas}
ds_k^2 \,\approx\, \frac{\Lambda}{3} \mathcal{R}^2 dt^2 - \frac{3}{\Lambda} \frac{d\mathcal{R}^2}{\mathcal{R}^2}
 + \mathcal{R}^2 \left( d\theta^2 + \sin^2 \theta d\varphi^2 \right) \;,\qquad |\mathcal{R}| \gg M 
\end{equation}
which is asymptotic behavior of de Sitter space. Both asymptotic regions $\mathcal{R} \to \pm\infty$ have the same de Sitter radius equal to $\ell=\sqrt{3/\Lambda}$. The region $r>M$ can be called an asymptotically de Sitter wormhole with a throat located at $r=2M$ ($R=0$) with a proper radius $2\sqrt{2}M$ and cosmological horizons located at $r=r_+$ and $r=r_{++}$.

As for the region $0<r<M$ a better choice for radial coordinate is 
$\mathcal{R} \equiv r/\Omega_{kc}(r)$ which uniformly goes from 0 to $\infty$ as $r$ goes from 0 to $M$. In the limit $r\to M$ ($\mathcal{R} \to \infty$) the metric (\ref{mtzkg}) behaves as in (\ref{mtzkgas}) which is asymptotically de Sitter with a radius $\ell$. In this region MTZ solution in k-gauge describes spherically symmetric asymptotically de Sitter naked singularity with cosmological horizon located at $r=r_-$.

In conclusion, we have showed that MTZ solution, when represented in a regular field redefinition scheme, is not a black hole at all, but instead describes collection of two spherically symmetric asymptotically de Sitter objects -- a wormhole and a naked singularity. Again, a singularity surface ($r=M$) is in fact a new asymptotic region from the perspective of regular schemes.

\vspace{10pt}

\section{Anabalon-Cisterna solution}
\label{sec:ac}

\medskip

In \cite{Anabalon:2012tu} it was shown that MTZ solution is a special case of a larger class of solutions, parametrized with an additional parameter $\xi$, corresponding to the Jordan frame action (\ref{ccsc}) with potential $V_c$ given by
\begin{equation} \label{acpot}
V_c(\chi) = - \frac{\kappa \Lambda}{36} \chi^4
 + \frac{\Lambda \sqrt{6\kappa}}{9} \frac{\xi}{\xi^2+1} \left(\frac{6}{\kappa} - \chi^2 \right) \chi
\end{equation}
The solutions are given by
\begin{eqnarray} \label{accg}
&& ds_c^2 \,=\, \frac{(r - (\xi+1) M)^2}{(r-M)^2}
 \left\{ - \left[(r-M)^{2} - \frac{\lambda}{3} r^4 \right] \frac{dt^2}{r^2} + \frac{r^2\, dr^2}{(r-M)^2 - \frac{\lambda}{3} r^4}
 + r^2 (d\theta^2 + \sin^2 \theta d\varphi^2) \right\}
\nonumber \\
&& \chi(r) \,=\, \sqrt{\frac{6}{\kappa}}\, \frac{(\xi+1)M - \xi r}{r - (\xi+1)M} \;, \qquad \qquad
 \lambda \,\equiv\, \Lambda \frac{(\xi^2 -1)^2}{\xi^2 +1}
\end{eqnarray}
and there is also the second branch obtained by applying $\chi \to -\chi$ and $\xi \to -\xi$ on (\ref{accg}) simultaneously.
For the sake of clarity of presentation let us restrict ourselves\footnote{The following analysis straightforwardly extends to other sectors of the parameter space.} to $|\xi|<1$, $\Lambda>0$, and $0 < M < L/4$, where $L \equiv \sqrt{3/\lambda}$. In this case the metric in (\ref{accg}) is asymptotically de Sitter as $r\to\infty$ with radius $L$, has a singularity at $r=0$, and three Killing horizons $r_{\pm}$ and $r_{++}$ defined by (\ref{mtzhor}) (with substitution $\ell \to L$) which satisfy $0 < r_- < M < r_+ < 2M < \frac{L}{2} < r_{++} < L$. At $r=(\xi+1)M$ scalar field $\chi$ is singular, while at $r=2M$ it takes value 
$\chi=\sqrt{6/\kappa}$. All these properties look rather similar to those obtained for MTZ black hole in Jordan frame (up to some ``deformations'' caused by $\xi \ne 0$). In addition, for $\xi\ne0$ in the limit $r\to M$ one has $R^{\mu\nu}{}_{\rho\sigma} = \frac{\lambda}{3\xi^2} \delta^{\mu\nu}_{\rho\sigma}$, which indicates that $r=M$ is also an asymptotic region with de Sitter radius equal to $\xi L$. It follows that (\ref{accg}) describes two separate asymptotically de Sitter configurations, one defined in $r<M$ (a naked singularity) and the other in $r>M$ (a wormhole-like).\footnote{For $\Lambda=0$ the only difference is that solutions are asymptotically flat instead of dS.} Note that we obtained this before for MTZ black hole, but \emph{only after} we passed to the some regular scheme (e.g., k-gauge). 

It is important to observe two additional singular properties of the solution (\ref{accg}):
\begin{itemize}
\item[(i)] For $\xi\ne0$ there is an additional space-time singularity at $r=(\xi+1)M$ (exactly where the scalar field $\chi$ is singular) appearing either in the wormhole (for $\xi>0$) or naked singularity (for $\xi<0$) subsolution.
\item[(ii)] The solution with $\xi=0$ has a metric which is well-defined across $r=M$, because of the cancellation of numerator and denominator of the factor in (\ref{accg}). This is in fact MTZ solution studied in the previous section. We have here an example in which maverick solution appears as a singularity in the continuous set of solutions (belonging to theories with different potentials).
\end{itemize}

Armed with the experience from previous sections, we can now easily understand what is happening here by using WIDG language. First, note that the action (\ref{ccsc}) with potential (\ref{acpot}) can be obtained from WIDG action (\ref{Iwism}) with a scalar potential
\begin{equation} \label{vac}
V(\phi,h) =  \frac{\kappa \Lambda}{36} (\phi^2 - h^2) \left( \phi^2 + h^2 + \frac{4\xi}{\xi^2 + 1} \phi h \right)
\end{equation}
by using c-gauge (\ref{cgauge}) and defining $\chi = h_c$. Let us follow the solution (\ref{accg}) in $\phi$-$h$ plane which at $r=\infty$ starts in the right wedge (see Fig. \ref{fig:gauges}), then passes $\phi =  h$ at $r=2M$ and enters upper wedge (antigravity region). After that behavior depends on the sign of $\xi$. For $\xi<0$ solution then reaches the second asymptotic region at $r=M$ where $h/\phi = 1/|\xi| > 0$ which guarantees that c-gauge is regular. As a consequence the solution in c-gauge for $r>M$ is regular. However, as we follow the c-gauge solution in $r<M$ region we hit the surface $r=(\xi+1)M$ at which both the metric and $h_c$ are singular. But we can now anticipate what is going on -- in regular gauges for $r=(\xi+1)M$ one hits $\phi=0$ which is a singular line for c-gauge. We conclude that there is no problem with the solution but with the gauge (field redefinition scheme) and if we use regular gauge this singularity should go away. Let us mention that if we took instead $\xi>0$ the only difference in the argument is that ``singularity'' $r=(\xi+1)M$ is on $r>M$ side (in wormhole solution).

We now proceed as before by taking k-gauge as a representative of regular gauges (for this situation). It is easy to show that for the potential (\ref{vac}) the action in k-gauge is 
\begin{equation} \label{acIk}
I_k = \frac{1}{2\kappa} \int d^4 x \sqrt{-g_k} \cos(2\beta) \left\{ R_k - 6 (\partial \beta)^2
 - 2 \Lambda \left[ 1 + \frac{2\xi}{\xi^2+1} \sin(2\beta) \right] \right\}
\end{equation}
We see that scalar potential in k-gauge is rather simple, and this may possibly explain why analytic solutions are obtainable.
The easiest way to obtain a solution in k-gauge is by Weyl-rescaling the c-gauge solution (\ref{accg}) by a factor $\Omega_{kc}$ calculated from (\ref{transom}). The result is
\begin{eqnarray} \label{ackg}
&& ds_k^2 \,=\, \frac{(\xi^2+1)r^2 - 2(\xi+1)^2 M(r-M)}{(r-M)^2}
 \left\{ - \left[(r-M)^{2} - \frac{\lambda}{3} r^4 \right] \frac{dt^2}{r^2} + \frac{r^2\, dr^2}{(r-M)^2 - \frac{\lambda}{3} r^4}
 + r^2 d\Omega_2 \right\}
\nonumber \\
&& \beta(r) \,=\, \arctan \frac{(\xi+1)M - \xi r}{r - (\xi+1)M}
\end{eqnarray}
A simple analysis reveals that solution in k-gauge consists of two asymptotically de Sitter pieces, $0<r<M$ part describes a naked singularity while $r>M$ part describes a wormhole with a throat located at $r=2M$ with a radius equal to $2\sqrt{2}(1-\xi)M$. de Sitter radia of asymptotic regions $r\to\infty$ and $r=2M$ are now equal and given by $L\sqrt{1+\xi^2}$. We emphasize the following results of our analysis of Anabalon-Cisterna solution:
\begin{itemize}
\item In k-gauge (as in all regular field redefinition schemes) there are no singularities aside the one in $r=0$.\footnote{By a proper choice of a scheme even this singularity can be regularized, see \cite{Prester:2013fia}. However, this is not essential for purposes of the present paper and we omit introducing this complication here.}.  For solutions which are regular in c-gauge the physical interpretation is the same in c-gauge (Jordan frame) and k-gauge.
\item In k-gauge solution for $\xi \ne 0$ (MTZ solution) is not special and has essentially the same physical interpretation as other solutions with $|\xi|<1$. 
\end{itemize}
These observations strengthen our claim that the proper physical interpretation of BBMB and MTZ solution are obtained in regular field redefinition schemes (such as k-gauge).

\vspace{10pt}

\section{Discussion and conclusion}
\label{sec:concl}

\medskip

We have analyzed BBMB solution, and its generalizations (such as MTZ solution), which are strange ("maverick") solutions with scalar hair in the theory with scalar field conformally coupled to gravity. The metric part of the solution is the same as for particular Reissner-Nordstom black hole, but the scalar field develops a singularity which may compromise the black hole interpretation. Our observation is that singularity appears exactly at the place where one expects break-down of field redefinition scheme (Jordan frame in this case) in which solution is originally constructed and presented. After passing to a regular field redefinition scheme, the solution becomes regular but the interpretation changes. Instead of describing black hole, BBMB/MTZ solution describes a collection of two solutions with different physical interpretations - a naked singularity and a wormhole. This is because singular surface is in fact a new asymptotic region when solutions are described in regular field redefinition schemes. Our conclusion is also supported by viewing BBMB/MZT solution as a special case of a broader class of solutions (constructed in \cite{Anabalon:2012tu}). In regular schemes an interpretation is the same for the larger class of solutions, contrary to the Jordan frame scheme where BBMB/MTZ solutions are exceptions (have singular interpretations). It is trivial to show that our conclusions extend to all known generalizations of these solutions, e.g., electrically charged and so on.

Embedding of the original Lagrangian into the framework of Weyl-invariant dilaton gravity (WIDG) gave us a perfect tool to study domain of configuration space and consequently to locate the source of a singularity to the breakdown of field redefinition scheme. WIDG formulation allowed us to construct a new scheme (which we nicked k-gauge) which accommodates larger configuration space then in Jordan or Einstein frame, in which solutions can be regularly represented. We believe that this scheme may find uses more broadly, e.g., in searches for new solutions, either in cosmological, black hole/wormhole or other contexts. We note that the theory for which analytic solutions are found in \cite{Anabalon:2012tu}, looks more simple when expressed in k-gauge frame instead of Jordan frame, and this could be part of the answer why closed-form analytic solutions were found at all for such not-trivial scalar potentials. It should be mentioned that WIDG formulation attracted some attention in recent years, e.g., in analyzes of new cyclic solutions \cite{Bars:2013yba} or inflationary models \cite{Kallosh:2013hoa} in cosmology, high-energy behavior of Standard Model \cite{tHooft:2011aa,Hooft:2010ac}, and black hole singularities in \cite{Prester:2013fia}. Here we emphasized its usefulness as a tool in analyzes of connections and differences between different field redefinition schemes, a topic which still occasionally leads to confusion \cite{Faraoni:2006fx}.

\vspace{25pt}






\end{document}